\documentclass[reprint,amssymb, amsmath, aps, superscriptaddress, showpacs, footinbib, prb]{revtex4-1}
\usepackage{graphicx}
\usepackage{color}
\usepackage[colorlinks,breaklinks,bookmarks=true,citecolor=blue,linkcolor=red,urlcolor=blue]{hyperref}
\newcommand{\eff}{\text{eff}}
\newcommand{\AFM}{\text{AFM}}

\newcommand{\Dv}{\mathbf D}

\begin{document}

\title{The spin gap in malachite Cu$_2$(OH)$_2$CO$_3$ and its evolution under pressure}

\author{Stefan Lebernegg}
\email{stefan.l@sbg.at}
\affiliation{Max Planck Institute for Chemical Physics of Solids, N\"{o}thnitzer
Str. 40, 01187 Dresden, Germany}

\author{Alexander A. Tsirlin}
\affiliation{Max Planck Institute for Chemical Physics of Solids, N\"{o}thnitzer
Str. 40, 01187 Dresden, Germany}
\affiliation{National Institute of Chemical Physics and Biophysics, 12618 Tallinn, Estonia}

\author{Oleg Janson}
\affiliation{Max Planck Institute for Chemical Physics of Solids, N\"{o}thnitzer
Str. 40, 01187 Dresden, Germany}
\affiliation{National Institute of Chemical Physics and Biophysics, 12618 Tallinn, Estonia}

\author{Helge Rosner}
\email{Helge.Rosner@cpfs.mpg.de}
\affiliation{Max Planck Institute for Chemical Physics of Solids, N\"{o}thnitzer
Str. 40, 01187 Dresden, Germany}
\date{\today}

\begin{abstract}
We report on the microscopic magnetic modeling of the spin-1/2 copper mineral malachite at ambient and elevated pressures. Despite the layered crystal structure of this mineral, the ambient-pressure susceptibility and magnetization data can be well described by an unfrustrated quasi-one-dimensional magnetic model. Weakly interacting antiferromagnetic alternating spin chains are responsible for a large spin gap of 120\,K. Although the intradimer Cu--O--Cu bridging \hbox{angles} are \hbox{considerably} smaller than the interdimer angles, density functional theory (DFT) calculations \hbox{revealed} that the largest exchange coupling of 190\,K operates within the structural dimers. The lack of the inversion symmetry in the exchange pathways gives rise to sizable Dzyaloshinskii-Moriya interactions which were estimated by full-relativistic DFT+$U$ calculations. Based on \hbox{available} high-pressure crystal structures, we investigate the exchange couplings under pressure and make \hbox{predictions} for the evolution of the spin gap. The calculations evidence that intradimer couplings are strongly pressure-dependent and their evolution underlies the decrease of the spin gap under pressure. Finally, we assess the accuracy of hydrogen positions determined by structural \hbox{relaxation} within DFT and put forward this computational method as a viable alternative to \hbox{elaborate} \hbox{experiments}.
\end{abstract}

\maketitle

\section{Introduction}

Cu-based minerals enjoy close attention of researchers working in the fields of
geology and solid-state physics alike. Intricate crystal structures underlie
complex arrangements of the spin-$\frac12$ Cu$^{2+}$ ions that, in turn,
trigger interesting low-temperature quantum effects and exotic ground
states.\cite{balents2010} For example, herbertsmithite Cu$_3$Zn(OH)$_6$Cl$_2$,
the best available spin-$\frac12$ kagom\'e system, shows putative spin-liquid
ground state.\cite{balents2010,[{}][{, and references
therein.}]mendels2010,helton2010,freedman2010,han2012} Dioptase
Cu$_6$Si$_6$O$_{18}\cdot 6$H$_2$O demonstrates unusually strong quantum
fluctuations on a non-frustrated three-dimensional spin lattice.\cite{[{}][{,
and references therein.}]dioptase} Azurite Cu$_3$(CO$_3)_2$(OH)$_2$ reveals a
$\frac13$-plateau in the magnetization and presumably hosts a rare magnetic
topology of the diamond spin
chain.\cite{kikuchi2005,rule2008,*gibson2010,*rule2011,aimo2009,*aimo2011,azurite}
Linarite PbCuSO$_4$(OH)$_2$ is an excellent material prototype of the
strongly frustrated spin chain.\cite{wolter2012,*yasui2011,willenberg2012}
Finally, volborthite Cu$_3$V$_2$O$_7$(OH)$_2\cdot$2H$_2$O that was originally considered as
a kagom\'e material\cite{hiroi2001,*fukaya2003,*yamashita2010,bert2005,yoshida2009,*yoshida2012}
reveals a more complex and still enigmatic frustrated spin
lattice.\cite{janson2010b,nilsen2011}

Malachite is arguably the best known Cu secondary mineral typically formed in
the oxidation zone of Cu deposits as weathering product of Cu sulphides. The
earliest source of copper (minerals quarried together with malachite were a
convenient flux that facilitated the smelting),\cite{fleming1993} it was
extensively used as ornamental stone and as a green
pigment\cite{bruni1999,*eremin2006,*burgio2010} since antiquity. The related
famous blue Cu-carbonate azurite transforms to malachite by absorption of water
and loss of CO$_2$. This transformation known as "greening" is responsible for
greenish instead of blue skies on some historical frescos.\cite{pigm} More
recently, malachite and its Zn-substituted versions were recognized as a
convenient precursor of mixed CuO--ZnO
catalysts.\cite{behrens2009,*behrens2010} 

Regarding this long and prominent history of malachite, surprisingly little is known about its magnetism.
Janod~\textit{et al.}\cite{janod2000} reported the sizable spin gap of about
130~K and proposed a one-dimensional model of bond-alternating spin chains.
This model emerges naturally from the crystal structure of
malachite,\cite{zigan1977} where CuO$_4$ plaquettes form Cu$_2$O$_6$ dimers by
edge-sharing and further link into chains along the $[201]$ direction by
corner-sharing. According to Janod~\textit{et~al.},\cite{janod2000} the
stronger coupling should run between the structural dimers because of the
larger Cu--O--Cu angle that promotes antiferromagnetic (AFM) superexchange.
This suggestion, inferred from the well-known Goodenough-Kanamori-Anderson (GKA)
rules,\cite{GKA1,GKA2,GKA3} is in line with many recent studies of
Cu$^{2+}$-based compounds, where structural Cu$_2$O$_6$ dimers do not match the
spin dimers and show weak magnetic couplings,
only.\cite{tsirlin2010,*cu2x2o7,mentre2009,*tsirlin2010b,benyahia2006,deisenhofer2006,clinoclase} 

Here, we present a detailed density functional theory (DFT)-based microscopic study of malachite and support the
ensuing model by magnetization measurements combined with quantum Monte-Carlo (QMC) simulations of thermodynamic properties.
Contrary to Janod~\textit{et al.},\cite{janod2000} we argue that malachite is a
rare Cu$^{2+}$ system where structural dimers match the spin dimers. We discuss
the origin of this effect, and provide a comprehensive picture of magnetic
exchange parameters, including both isotropic and anisotropic exchange
couplings. Eventually, we take advantage of available high-pressure crystal
structures of malachite\cite{malachiteP} and investigate the pressure
dependence of the exchange couplings and the ensuing spin gap.

The paper is organized as follows. In Sec.~\ref{sec:methods}, the applied experimental and
theoretical methods are presented. The crystal structure of malachite is
described in Sec.~\ref{sec:xxstr}. The experimental and computational results for the
ambient-pressure malachite are provided in Sec.~\ref{sec:zeroP}. Our predictions for the
evolution of the spin gap under pressure are presented in Sec.~\ref{sec:highP}. Finally, 
discussion and summary are given in Secs.~\ref{sec:discussion} and~\ref{sec:summary}, respectively.

\section{Methods}
\label{sec:methods}
For our experimental studies we used a natural sample of needle-shaped
malachite from Tsumeb, Namibia. The sample
was investigated by laboratory powder x-ray diffraction (XRD) (Huber G670
Guinier camera, CuK$_{\alpha\,1}$ radiation, ImagePlate detector,
$2\theta\,=\,3-100^{\circ}$ angle range). High-resolution low-temperature XRD
data were collected at the ID31 beamline of the European Synchrotron Radiation
Facility (ESRF, Grenoble) at a wavelength of about 0.35\,\r{A}. The chemical
composition was determined by the ICP-OES method.\footnote{ICP-OES (inductively
coupled plasma optical emission spectrometry) analysis was performed with the
Vista instrument from Varian.} The magnetization was measured with a Quantum Design
MPMS SQUID magnetometer in a temperature range of 2-400\,K in fields up to
5\,T. For measurements up to 14\,T, a vibrating sample magnetometer setup of a Quantum Design PPMS was used.

Electronic and magnetic structure calculations were performed within DFT by using the full-potential local-orbital code
\texttt{fplo9.07-41}.\cite{fplo} Local density (LDA)\cite{pw92} and generalized
gradient approximations (GGA)\cite{pbe96} were used for the
exchange-correlation potential together with a well converged $k$-mesh of
5$\times$5$\times$5 points for the crystallographic unit cell and about 100 points for supercells. For the
optimization of hydrogen positions, in addition to \textsc{fplo}, the Vienna Ab initio Simulation Package (\texttt{VASP5.2})\cite{vasp1,*vasp2} was used in combination with LDA, GGA,
revPBE,\cite{revPBE} DFT-D\cite{dft_d} and HSE06\cite{hse06} exchange-correlation functionals. For
a full relaxation of all atomic positions in the high-pressure structures, we
employed the GGA+$U$ method implemented in \texttt{VASP}. 

Strong electronic correlations were included in two different ways: First, by
mapping the LDA bands onto an effective one-orbital tight-binding (TB) model.
Thereby, the transfer integrals $t_{ij}$ of the TB-model are evaluated as
nondiagonal matrix elements between Wannier functions (WFs). These transfer
integrals $t_{ij}$ are further introduced into the half-filled single-band Hubbard
model
$\hat{H}=\hat{H}_{TB}+U_{\text{eff}}\sum_{i}\hat{n}_{i\uparrow}\hat{n}_{i\downarrow}$, with $U_{\eff}$ being the effective onsite Coulomb repulsion.
In case of half filling and for the strongly correlated limit $t_{ij}\ll
U_{\eff}$, as realized in malachite (see Table~\ref{tJ}), the Hubbard model can be reduced to the Heisenberg model for the low-energy excitations, 
\begin{equation} 
\hat{H}=\sum_{\left\langle ij\right\rangle}J_{ij}\hat{S_{i}}\cdot\hat{S_{j}},
\label{eq:heis}\end{equation}
where $\left\langle ij\right\rangle$ denotes the summation over all bonds of
the spin lattice. Accordingly, the antiferromagnetic (AFM) contributions to the
exchange coupling constants can be evaluated in second order perturbation theory as 
$J_{ij}^{\AFM}=4t_{ij}^2/U_{\eff}$.

Alternatively, the full exchange couplings $J_{ij}$, comprising ferromagnetic (FM)
and AFM contributions, can be derived from total energy differences of various
collinear spin arrangements evaluated in spin-polarized supercell calculations
within the mean-field DFT+$U$ formalism. For
the double counting correction, a fully localized limit approximation was used and the
on-site Coulomb repulsion and onsite Hund's exchange of the Cu(3$d$) orbitals
are chosen as $U_d$\,=\,8.0$\pm$1.0\,eV and $J_d$\,=\,1.0\,eV, respectively,
similar to parameter sets we have used previously for other
cuprates.\cite{azurite,dioptase}

The anisotropic exchange was calculated with the full relativistic version of
GGA+$U$ provided by \texttt{VASP} with $U_d$\,=\,9.5\,eV, $J_d$\,=\,1.0\,eV and
the default projector-augmented wave (``PAW-PBE'')
pseudopotentials\cite{vasp_pseudo} on a 4$\times$4$\times$4 $k$-mesh. For each
exchange ($J$ and $J'$) 36 magnetic configurations (four configurations for
each matrix element of the exchange matrix) were calculated.\cite{dm_cal}
The $U_d$ parameter of GGA+$U$ was chosen so that the isotropic exchanges
$J_{ij}$ obtained from \texttt{VASP} agree with those from the \texttt{FPLO}
calculations. The 1.5\,eV offset in the $U_d$ values arises from the different
exchange-correlation functionals and different basis sets used by the two
codes.

Quantum Monte Carlo (QMC) simulations were performed using the codes
\textsc{loop}\cite{loop} and \textsc{dirloop\_sse}\cite{dirloop} from the
software package \textsc{alps-1.3}.\cite{ALPS} Magnetic susceptibility and
magnetization of the two-dimensional model were simulated on finite lattices
comprising up to $N$\,=\,1024 spins, using periodic boundary conditions. For
simulations in zero field, we used 200\,000 sweeps for thermalization and
2\,000\,000 sweeps after thermalization. For finite-field simulations, 40\,000
and 400\,000 sweeps were used, respectively.

\section{Crystal structure}
\label{sec:xxstr}
\begin{figure}[tb] \includegraphics[width=8.6cm]{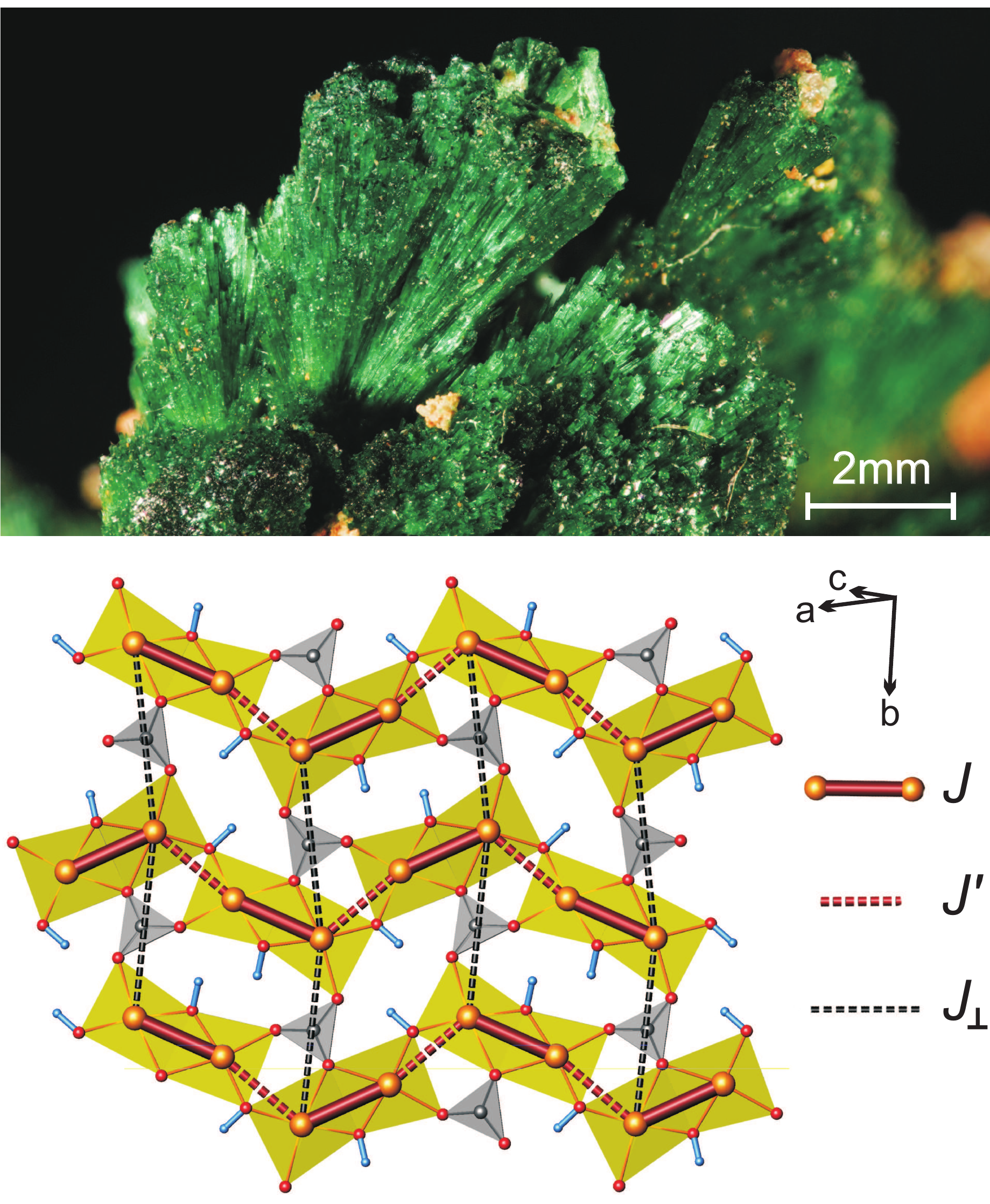}
\caption{\label{xxstr}(Color online) Natural malachite specimen from Tsumeb, Namibia, with aggregates of needle-shaped crystals (top). Bottom: The ambient pressure crystal structure and the microscopic magnetic model of malachite. The structural sketch shows Cu$_2$O$_6$ dimers (yellow), CO$_3$ triangles (grey), and OH bonds (blue). Leading magnetic exchange couplings are shown as thick cylinders (legend in the right panel). Alternating spin chains are
formed by $J$ and $J'$.
}
\end{figure}

Malachite crystallizes in the monoclinic space group $P2_1/a$ with the lattice
constants $a$\,=\,9.5020\,\r{A}, $b$\,=\,11.9740\,\r{A}, $c$\,=\,3.240\,\r{A}
and the monoclinic angle $\beta$\,=\,98.75$^{\circ}$.\cite{zigan1977}
Nearly planar CuO$_4$ plaquettes form doubly bridged Cu$_2$O$_6$ dimers by edge-sharing (Fig.~\ref{xxstr}). The dimers themselves share common corners and form slightly twisted chains running along the [201] direction.
The Cu--O--Cu bridging angles within the dimers are rather different with 94.7$^{\circ}$ and 106.4$^{\circ}$, respectively, resulting in an average bridging angle of 100.5$^{\circ}$. The angle between the dimers amounts to 122.1$^{\circ}$. Carbonate groups
link the chains to planes parallel ($\overline{2}$01) which are separated from each other by about 2.37\,\r{A}~\cite{supplement} and are responsible for the
perfect cleavage of malachite. Owing to the short interplane distance, however, the Mohs hardness of this mineral, 3.5--4, is significantly higher than for other
layer-structured Cu-minerals as, e.g., clinoclase (2.5--3) or posnjakite (2--3).\cite{mohs} There is another fair cleavage on (010), i.e. orthogonal to planes and parallel to chains, that
breaks the bonds between dimers and the CO$_3$ groups.

Malachite represents one of the rare cases of Cu-minerals for which accurate hydrogen positions are available from the experiments. The positions of hydrogen atoms are essential for deriving reliable microscopic magnetic models. Though H is not directly involved in the magnetic superexchange process, in particular its bonding angle has a crucial effect on the type and strength of the exchange couplings.\cite{ruiz97_2,clinoclase} The standard experimental technique for determining H-positions is neutron scattering which is, however, difficult, expensive and requires big single crystals ($>1$\,mm$^3$), preferably of deuterium-enriched samples. For these reasons, H-positions remain undetermined for most minerals. Alternatively, the hydrogen positions could be obtained by structural optimization within DFT. However, to establish such a procedure, first a careful analysis and comparison with experimental data is mandatory. Malachite, thus, provides an excellent opportunity to test several different DFT functionals and compare the results with the neutron data (Table~\ref{H_opt}). 

Standard LDA and GGA functionals should provide a reasonable description of the covalent O--H bonds. However, they may have deficiencies in describing weak dispersion effects,\cite{dft_d} in particular longer O$\cdots$H hydrogen bonds that determine the direction of the covalent O--H bond. Since there is no functional particularly designed for an accurate description of hydrogen bonds, we test different functionals that either incorporate empirical corrections for van der Waals interactions (DFT-D) or contain different adjustments providing an improved description of lattice parameters, cohesive energies, and different bond lengths (PBEsol, revPBE, HSE06).

Although footing on very different levels of sophistication, all tested functionals supply excellent results, in particular, for the relevant bond angles (Table~\ref{H_opt}). The bond lengths are generally overestimated, with the typical deviation of $1-2$\,\%, the largest deviation of 3\,\% for LDA, and the smallest deviation of less than 0.5\,\% for HSE06. It thus seems that HSE06 provides the best results. However, the structural data for malachite are obtained at room temperature, whereas DFT results pertain to the crystal structure at 0\,K. The experimental bond lengths are thus longer because of thermal expansion. Additionally, room temperature data are affected by the libration, a strong rocking vibration of the O--H bond that shortens the apparent O--H distance. This effect is well seen in the oblate thermal ellipsoids of hydrogen atoms that are stretched in the directions transverse to the O--H bonds.\cite{zigan1977} These competing but unquantified effects prevent a clear distinction of a certain functional.
However, the deviations between theory and experiment shown in Table~\ref{H_opt} are quite small and demonstrate that any functional can be used for optimizing hydrogen positions without significant loss of accuracy. These results show the capability of DFT to provide accurate hydrogen positions in Cu$^{2+}$ minerals and, potentially, in other hydrogen-containing compounds. DFT, therefore, represents a valuable alternative to elaborate experiments. The reliable determination of hydrogen positions further enables us to calculate the missing H-position in the high-pressure structures of malachite\cite{malachiteP} and to assess the magnetic behavior of malachite under pressure.

\begin{table*}[tbp]
\begin{ruledtabular}
\caption{\label{H_opt} 
Distances (\r{A}) and angles (deg) of hydrogen obtained by structural
optimization using the \texttt{fplo} and \texttt{VASP} codes in combination
with various exchange-correlation functionals. "dev" is the averaged deviation (in \%) between theoretical and experimental\cite{zigan1977} data
for distances and angles, respectively.}
\begin{tabular}{*9c}
  & \multicolumn{5}{c}{\texttt{VASP}} & \multicolumn{2}{c}{\texttt{fplo}} & \\ \cline{2-6} \cline{7-8}
            & PBE    & PBEsol & revPBE & DFT-D & HSE06  & LDA    & PBE     & experiment \\ \hline
H1-O4       & 0.995  & 1.002  & 0.991  & 0.996 & 0.976  & 1.006  & 0.995   & 0.975(2) \\ 
H2-O5       & 0.991  & 0.998  & 0.987  & 0.991 & 0.972  & 1.004  & 0.992   & 0.969(2) \\ 
H2-H2       & 1.916  & 1.932  & 1.912  & 1.918 & 1.899  & 1.943  & 1.918   & 1.892(2) \\
dev         & 1.9    & 2.6    & 1.5    & 1.9   & 0.3    & 3.2    & 1.9     & \\
\\ 
H1-O4-Cu1   & 115.1 & 115.8  &  114.9  & 115.2 & 115.7  & 115.90 & 114.92 & 115.21(13) \\ 
H1-O4-O2    & 157.9 & 158.5  & 157.8   & 158.3 & 158.7  & 158.45 & 157.62 & 158.17(11) \\ 
H2-O5-Cu2   & 104.1 & 103.5  & 104.2   & 104.4 & 104.7  & 103.27 & 104.00 & 104.99(14) \\
H2-O5-O2    & 140.9 & 140.5  & 140.6   & 141.7 & 140.3  & 140.75 & 140.68 & 140.64(16) \\
dev         & 0.3   & 0.6    & 0.3     & 0.4   & 0.3    & 0.6    & 0.4    & \\
\end{tabular}
\end{ruledtabular}
\end{table*}

\section{Malachite at ambient pressure}
\label{sec:zeroP}

\subsection{\label{sample}Sample characterization}

Powder XRD measurements confirmed the purity of our malachite sample. The high sample quality was additionally supported by chemical analysis yielding 56.8(1)\% Cu, 0.1\% Pb, 0.1\% Ca and $<0.1$\% S. All other detectable elements, including
transition metals, are below 0.03\%. The lead impurity most likely stems from trace amounts of cerussite (PbCO$_3$), whereas sulphur may be present as elementary sulphur. Both PbCO$_3$ and S are found in the specimen matrix. Calcium may be attributed to calcite (CaCO$_3$). The low content of these formally nonmagnetic phases should have no disturbing effect on the magnetization measurements. 

The slight deficiency of Cu with respect to the expected value of 57.5\% may reflect trace amounts of amorphous impurities, which are not seen in XRD but affect the overall composition measured by chemical analysis. Our high-resolution XRD data and the ensuing structure refinement\cite{supplement} rule out any Cu deficiency in the crystalline phase of malachite. Additionally, we performed a structure refinement at 80\,K and confirmed that the structure does not change upon cooling. Our data perfectly match earlier room-temperature neutron data.\cite{zigan1977}

\subsection{\label{S-magn}Thermodynamical measurements}
\begin{figure}[tbp]
\includegraphics[width=8.6cm]{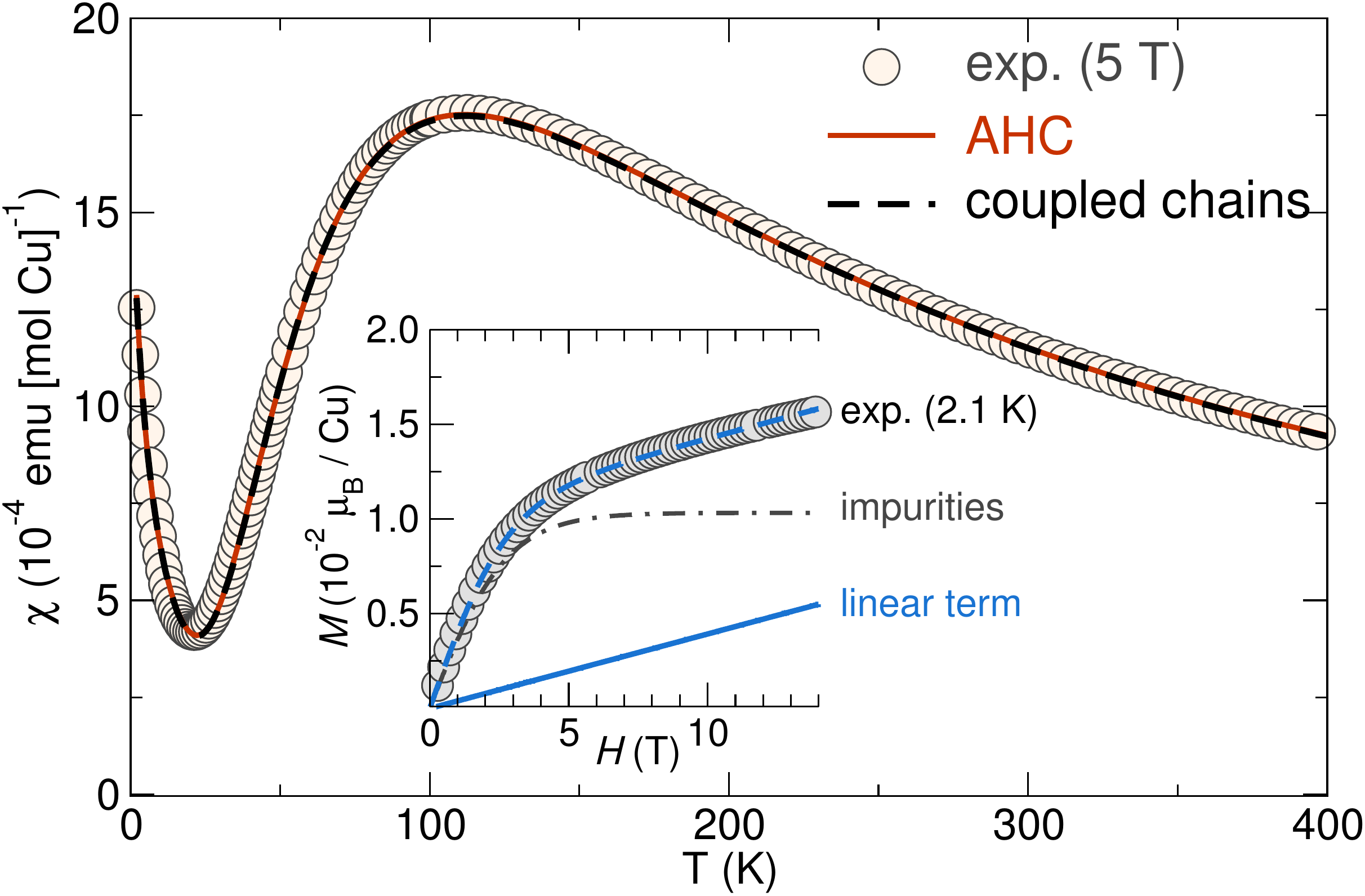} \caption{\label{F-exp} (Color
online) Magnetic susceptibility of malachite at ambient pressure (circles).
The solid line shows the parameterized solution of the alternating chain model
(AHC), adopted from Ref.~\onlinecite{johnston2000}. The dashed line is the
intrinsic contribution of the magnetic planes, as yielded by QMC simulations
for the full microscopic model. Inset: experimental magnetization isotherm
(circles) and its fit (dashed line) using two terms: paramagnetic impurity
contribution (dash-dotted line), described by the Brillouin function, and a
linear term (solid line).
}
\end{figure}

The temperature dependence of the magnetic susceptibility $\chi(T)$ shows a
paramagnetic behavior at high temperatures and no sign of long-range magnetic
ordering down to 2\,K (Fig.~\ref{F-exp}). The Curie-Weiss fit,
$\chi(T)=\chi_0+C/(T+\theta)$, to the high-temperature part ($T$\,$>$\,250\,K) yields
$\chi_0$\,=\,$5\times10^{-6}$\,emu\,(mol Cu)$^{-1}$,
$C$\,=\,0.4802\,emu\,K\,(mol Cu)$^{-1}$ and $\theta$\,=\,120.5\,K. The
resulting effective magnetic moment
$\mu_{\text{eff}}$\,=\,1.960\,$\mu_{\text{B}}$ exceeds the spin-only value of
$S$\,=\,$\frac12$ ($\mu_{\text{eff}}\!\simeq$\,1.73\,$\mu_{\text{B}}$),
yielding the $g$-factor of 2.26, which is within the typical range for
Cu$^{2+}$ compounds.\cite{diaboleite,dioptase} The positive value of the Weiss
temperature $\theta$ and the broad maximum in the magnetic susceptibility
around 111\,K evidence that leading exchange couplings and spin correlations in
malachite are AFM.

Following Ref.~\onlinecite{janod2000}, we fit the experimental curve with the parameterized solution for an alternating Heisenberg chain\cite{johnston2000}
(Fig.~\ref{F-exp}) and obtain leading exchange couplings of 189\,K and 89\,K,
and the $g$-factor of 2.18, in agreement with Ref.~\onlinecite{janod2000}. The
spin gap can be estimated as:\cite{barnes1999}

\begin{equation}
    \Delta \approx J (1 - \alpha)^{\frac34} (1 + \alpha)^{\frac14}
\end{equation}
where $\alpha=J'/J$. In this way, we obtain a large spin gap of $\Delta$\,=\,129\,K. 
The low-temperature upturn can be reproduced by a paramagnetic impurity with
$C_{\text{imp}}$\,=\,0.00808\,emu\,K\,(mol Cu)$^{-1}$ which corresponds to a fraction of about 2\% spin-1/2 impurities. The temperature-independent contribution is
$3.20\times10^{-6}$ emu\,(mol Cu)$^{-1}$.

Despite the excellent agreement of the alternating chain model with the
experimental $\chi(T)$, small interchain couplings are inevitably present in
malachite. Although these couplings do not suffice to stabilize a long-range
order (LRO), they can substantially affect the value of the spin gap. To
understand the nature of interchain couplings, we evaluate a DFT-based
microscopic magnetic model.

\subsection{Microscopic magnetic model}
\label{sec:mag_mod}

\begin{figure}[tbp]
\includegraphics[width=8.6cm]{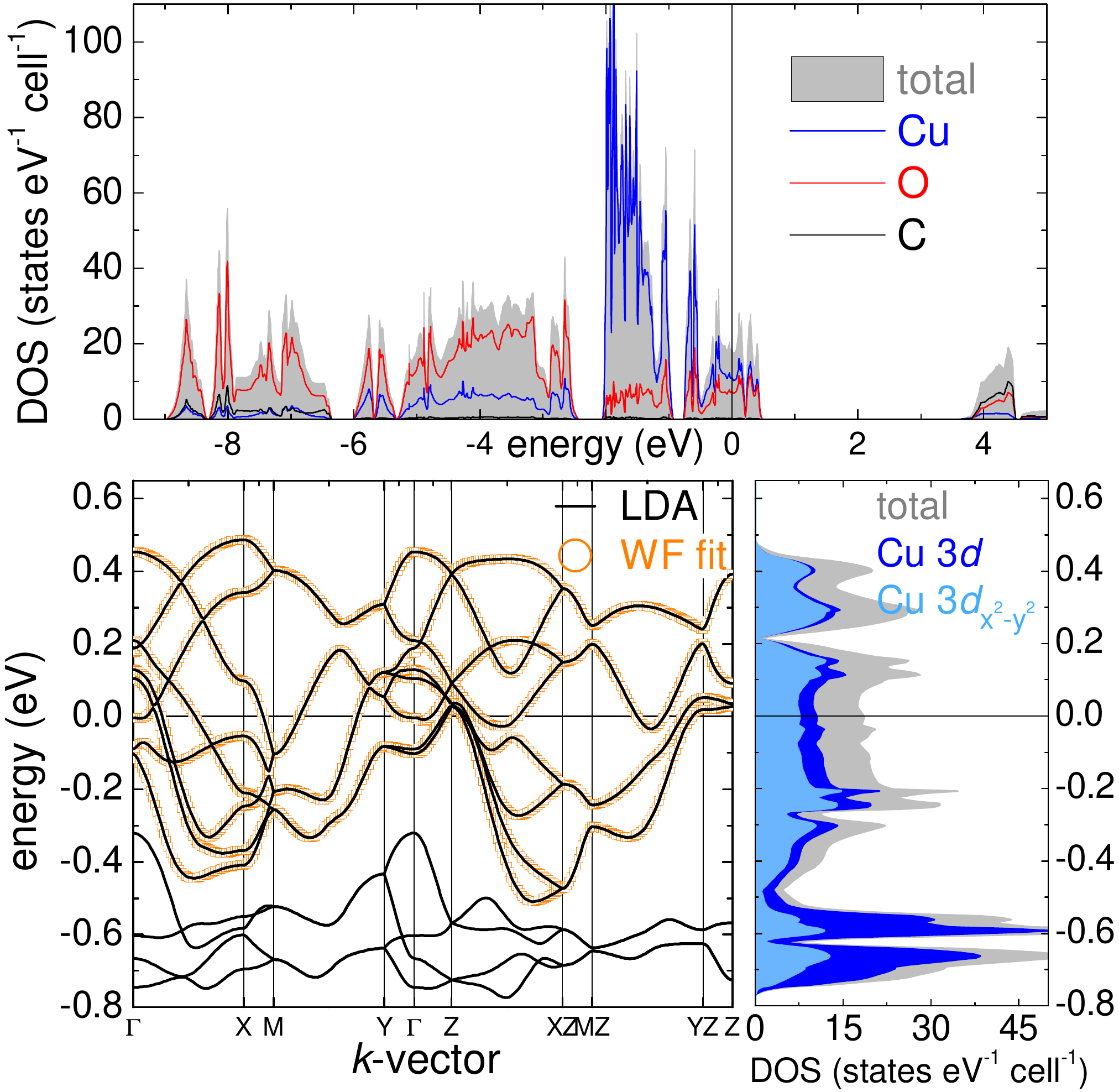}
\caption{\label{bands}
(Color online) Density of states (DOS) and the LDA valence bands of malachite.
The top panel shows the contributions from the Cu(3$d$), O(2$p$) and C valence
states to the total DOS. The Fermi level is at zero energy. In the bottom left
panel, the LDA-bands about the Fermi level are displayed and compared with bands
derived from a fit using an effective one-band tight-binding model based on
Cu-centered Wannier functions (WFs) projected on local
Cu(3$d_{x^2-y^2}$)-orbitals. The bottom right panel shows that the partial
Cu(3$d$)-DOS at the Fermi level is basically of Cu(3$d_{x^2-y^2}$) character,
justifying our construction of the WFs. }
\end{figure}

As a first step we performed LDA calculations that provide valuable information
about the crucial exchange pathways, though yielding a wrong metallic ground
state. The width of the LDA valence band block of about 9.5\,eV is typical for
cuprates (see Fig.~\ref{bands}). The blocks between $-2$\,eV and $-1$\,eV and between $-0.8$\,eV and $-0.5$\,eV, with a sizable Cu(3$d$) character, belong to antibonding $pd\pi^{\ast}$
and $pd\sigma^{\ast}$ orbitals, respectively. The $pd\sigma^{\ast}$ block can
be separated into bands with dominating Cu(3$d_{z^2-r^2}$) character, between $-0.8$\,eV and $-0.4$\,eV, and partially filled bands, crossing the Fermi level, with
Cu(3$d_{x^2-y^2}$) character. The orbitals are defined with respect to a local
coordinate system with the $x$-axis parallel to one of the Cu-O bonds and the
$z$-axis orthogonal to the CuO$_4$ plaquette plane. The small separation
between the two types of $pd\sigma^{\ast}$ bands arises from the relatively
short distance of about 2.37\,\r{A} between Cu2 and the apical oxygen, which
lifts the energy of the Cu(3$d_{z^2-r^2}$) orbital. 

The partially filled bands are sufficient to describe the low-lying magnetic
excitations and exchange couplings. Their projection onto a TB and subsequently
onto a Hubbard model yields the transfer integrals $t_{ij}$ and the corresponding
AFM contributions to $J_{ij}$ as given in Table~\ref{tJ}. Based on the GKA rules,
basically describing the dependency of the $t_{ij}$ on the bridging angle,
Janod~\textit{et~al.}\cite{janod2000} expected the $J$ coupling to be smaller
than $J'$ according to the smaller bridging angles of 94.7$^{\circ}$ and
106.4$^{\circ}$ for $J$ versus 122.1$^{\circ}$ for $J'$. Our results, however,
show exactly the opposite, with $J^{\AFM}$ being twice as large as $J'^{\AFM}$.
There are two reasons for this behavior. First, the sizable intradimer transfer
$t$ is indeed not unusual and can also be found in several other Cu-compounds
featuring doubly bridged Cu$_2$O$_6$ dimers. For example, in clinoclase the
respective transfer amounts to 191\,meV (101.9$^{\circ}$),\cite{clinoclase} in
Cu$_2$As$_2$O$_7$ $t=170$\,meV (101.7$^{\circ}$)~\cite{cu2as2o7} and in
Li$_3$CuB$_3$O$_7$ $t$ is 235\,meV (100.4$^{\circ}$).\cite{li3cub3o7} The
values in parentheses are the average intradimer bridging angles, which are
very similar to that in malachite (100.5$^{\circ}$). 

\begin{figure}[tbp]
\includegraphics[width=8.6cm]{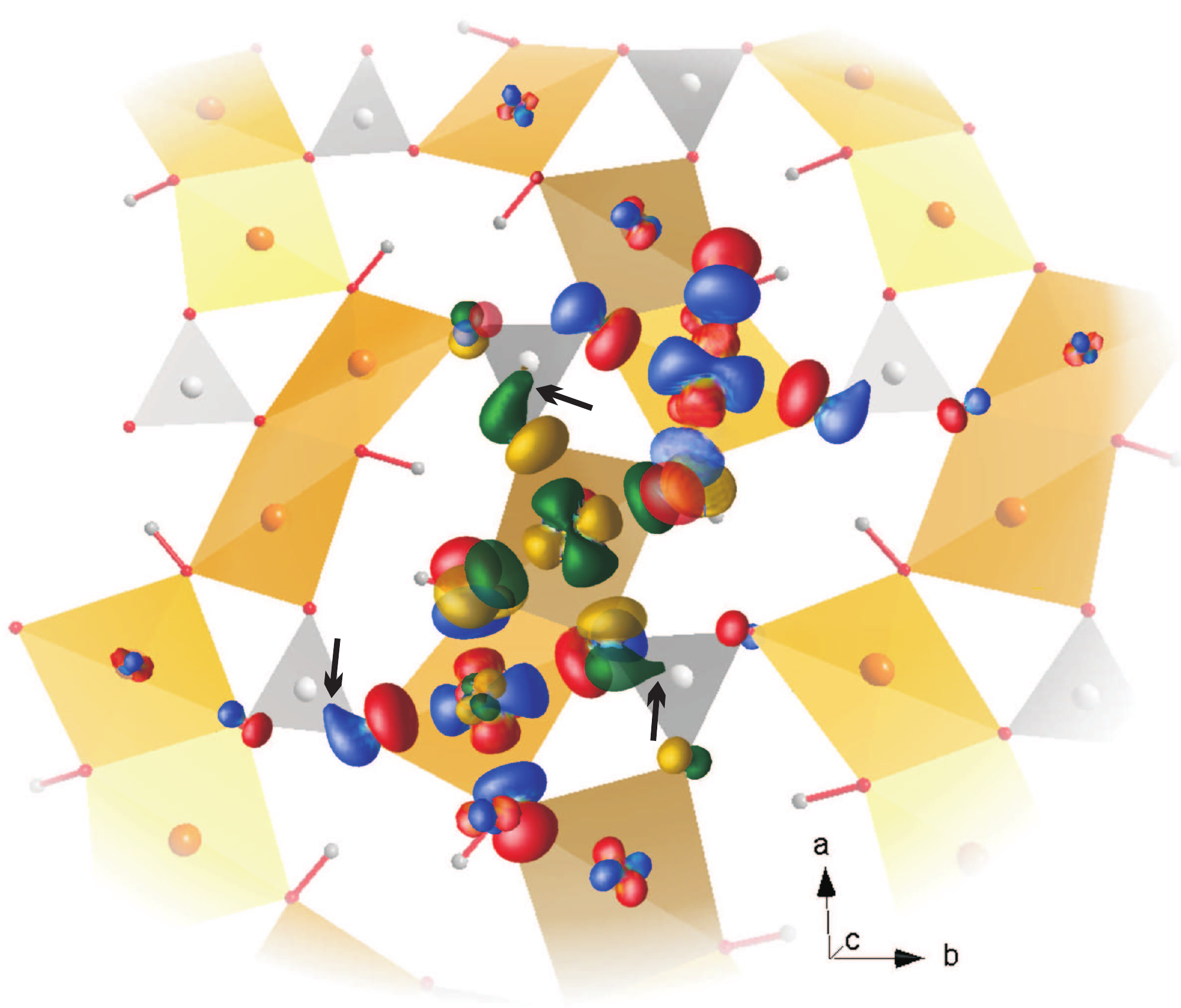}
\caption{\label{wf1}
(Color online) Wannier functions on the Cu1 (yellow-green) and Cu2 sites (red-blue). The polarization (marked by arrows) by the CO$_3$-groups is responsible for the large $t_{\perp}$ transfer integral. }
\end{figure}

The bridging angles alone do not explain why $J'^{\AFM}$ is so much smaller than $J^{\AFM}$ while featuring a much larger Cu--O--Cu bridging angle. Thus, second, the twisting of the chains has to be considered. The planes of the neighboring dimers are rotated against each other by about 20$^{\circ}$. This reduces the overlap of the WFs and, thus, the transfer integral $t'$. A similar effect of out-of-plane angles was discussed for clinoclase.\cite{clinoclase}
The fact that the long-range interchain transfer $t_{\perp}$ is of similar strength as $t'$ can, on one hand, be attributed to this reduction of $t'$ and, on the other hand, to the polarization of the WFs by the carbonate group (see Fig.~\ref{wf1}) that might significantly increase the overlap of the WFs despite the long Cu-Cu distance (1.8 times longer than that for $t'$). 
The short interlayer distance in malachite is responsible for a number of further sizable transfer integrals between the planes,\cite{supplement} which are, however, strongly reduced by FM contributions. All interlayer couplings are below 10~K and, thus, play a minor role for the microscopic magnetic model. 

\begin{table}[tbp]
\begin{ruledtabular}
\caption{\label{tJ} 
Transfer integrals $t_{ij}$ (in meV) and the AFM exchange contributions $J^{\text{AFM}}_{ij}=4t_{ij}^2/U_{\text{eff}}$ (in K) for $U_{\text{eff}}$\,=\,4.5\,eV. The total exchange couplings $J_{ij}$ are calculated with the LSDA+$U$ method and the parameters $U_d=8.0\pm1.0$\,eV, $J_d=1.0$\,eV. }
\begin{tabular}{c c c c c}
   &  Cu-Cu distance (\r{A}) & $t_{ij}$ & $J^{\AFM}_{ij}$ & $J_{ij}$  \\ \hline
$J$     & 3.06 & --143 & 211 & $193\pm40$  \\ 
$J'$ & 3.34 & --99 & 101 & $109\pm20$  \\
$J_{\perp}$ & 6.03 & --97 & 96 & $50\pm10$  \\ 
\end{tabular}
\end{ruledtabular}
\end{table}

The leading total exchange couplings $J_{ij}$, including AFM as well as FM contributions, are
calculated with the LSDA+$U$ method yielding the correct insulating ground
state of malachite. According to the small bridging angles, one could expect a
sizable reduction of $J$ due to FM contributions. However, $J^{\text{FM}}$
estimated as $J^{\text{FM}}=J-J^{\text{AFM}}$, does not exceed 60\,K so that
$J$ is still 1.5 times larger than $J'$ and, thus, remains the strongest
coupling in this compound. $J_{\perp}$ running via a
carbonate group, exhibits sizable FM contributions which is very
unusual since the FM contributions to long-range couplings are in general
small. However, similar effects have also been observed for the Cu-carbonate
azurite\cite{azurite} and CuSe$_2$O$_5$,\cite{cuse2o5} the latter featuring
pyramidal SeO$_3$ groups. By contrast, for exchange couplings
transmitted by tetrahedral PO$_4$ and AsO$_4$ ligands FM contributions play
a minor role.\cite{sr2cupo42,clinoclase} This interesting effect deserves a
closer look, which is, however, beyond the scope of the present work but will be
pursued in future studies. 

Now, we are at the position to simulate the temperature dependence of the
magnetic susceptibility for the full model. QMC yields the temperature
dependence of the reduced magnetic susceptibility $\chi^{\ast}$, which is scaled by adjusting
the overall energy $J$, the $g$-factor as well as extrinsic contributions:

\begin{equation}
\chi(T)=
\frac{N_Ag^2{\mu}_{B}^2}{k_B\,J}\cdot\chi^{*}\biggl(\frac{T}{k_B\,J}\biggl)
+ \frac{C_{\text{imp}}}{T+\theta_{\text{imp}}}+\chi_0.
\end{equation}

Our attempts to get a good fit by fixing the LSDA+$U$-based estimates
$J'/J$\,=\,0.56 and $J_{\perp}/J$\,=\,0.26 were not successful. To improve the
agreement, we varied the ratios in a reasonable range. The best fit can be
obtained by taking $J'/J$\,=\,0.45, $J_{\perp}/J$\,=\,0.15, which yields the
overall energy scale $J$\,=\,191\,K and the $g$-factor of 2.21 (the intrinsic
contribution is depicted as dashed line in Fig.~\ref{F-exp}).

The absolute value of $J$ perfectly agrees with the LSDA+$U$ estimate and also
nearly coincides with the estimate from the alternating chain fit. The second
largest exchange, $J'$, is somewhat smaller than the value supplied by LSDA+$U$
while the interchain coupling $J_{\perp}$ is substantially smaller with the
difference exceeding typical error bars. We argue that it could represent a
general tendency of DFT to overestimate the superexchange via CO$_3$ groups.
For instance, in the closely related compound, the natural mineral azurite
Cu$_3$(CO$_3$)$_2$(OH)$_2$, DFT also largely overestimates the interchain
coupling running though a CO$_3$ group.\cite{kang2009,azurite} Further
studies should shed light on this issue as we have already stressed before. 

The low-temperature part of the intrinsic magnetic susceptibility conforms to
an activated behavior:

\begin{equation} 
\chi \propto \exp\left(-\frac{\Delta}{T}\right)
\end{equation}

A fit of the intrinsic $\chi(T)$ up to 23\,K yields the spin gap $\Delta$\,=\,119\,K. Compared with the one-dimensional
alternating chain model (Sec.~\ref{S-magn}), yielding $\Delta$\,=\, 129\,K, the
interchain couplings reduce the spin gap in malachite by about 8\%.

\subsection{Dzyaloshinskii-Moriya couplings}
\label{sec:DM}
Since malachite features a large spin gap of $\Delta$\,=\,119\,K, its low-field
uniform magnetization is expected to be zero up to the critical field
$H_{c}\!=\!(g\cdot\mu_{\text{B}})^{-1}\Delta\!\simeq\!80$\,T, where the spin gap is
closed. In contrast, the experimental magnetization isotherm reveals that the
magnetization is non-zero and grows at least up to 14\,T (Fig.~\ref{F-exp},
inset). We also attempted a pulsed-field measurement in higher fields up to
60\,T, but no visible signal could be detected. This confirms that $H_{c}$
lies above 60\,T, whereas the signal below $H_{c}$ remains quite weak
($<3.5\times10^{-2}$\,$\mu_{B}$/Cu) and stays below the sensitivity limit of
our pulsed-field experiment. However, in a static-field experiment this weak
non-zero signal can be detected.
 
The low-field range is typically affected by paramagnetic impurities,
whose behavior is described by the Brillouin function. We can reproduce
the experimental data by the sum of the Brillouin function (the impurity
contribution is 1.8\%, its $g$-factor is 2.29, and the measurement
temperature is $T$\,=\,2.1\,K) and a linear $M=\gamma H$ term
(Fig.~\ref{F-exp}, inset). Although the temperature-independent
contribution $\chi_0$ would also lead to a linear increase in $M$, its
magnitude is way too small to explain our data: compare
$\gamma=3.9\times 10^{-4}$~$\mu_B$/T to the slope related to $\chi_0$,
which is on the order of 10$^{-5}$~$\mu_B$/T. Therefore, the nearly
linear growth of magnetization above 5\,T is likely of intrinsic
origin.

This experimental behavior can indicate the presence of anisotropic
Dzyaloshinskii-Moriya couplings.\cite{dm1,dm2} Such couplings break the SU(2)
invariance of the Heisenberg Hamiltonian [Eq.~\eqref{eq:heis}] and in the simplest case of an isolated dimer mix the singlet state
$\frac{1}{\sqrt{2}}(\mid\uparrow\downarrow\rangle -
\mid\downarrow\uparrow\rangle)$ with the zero-momentum triplet component
$\frac{1}{\sqrt{2}}(\mid\uparrow\downarrow\rangle +
\mid\downarrow\uparrow\rangle)$. In magnetic field, this mixing gives rise to
a finite magnetization and yields a linearly increasing
$M(H)$.\cite{miyahara2007}

\begin{figure}[tbp]
\includegraphics[width=8.6cm]{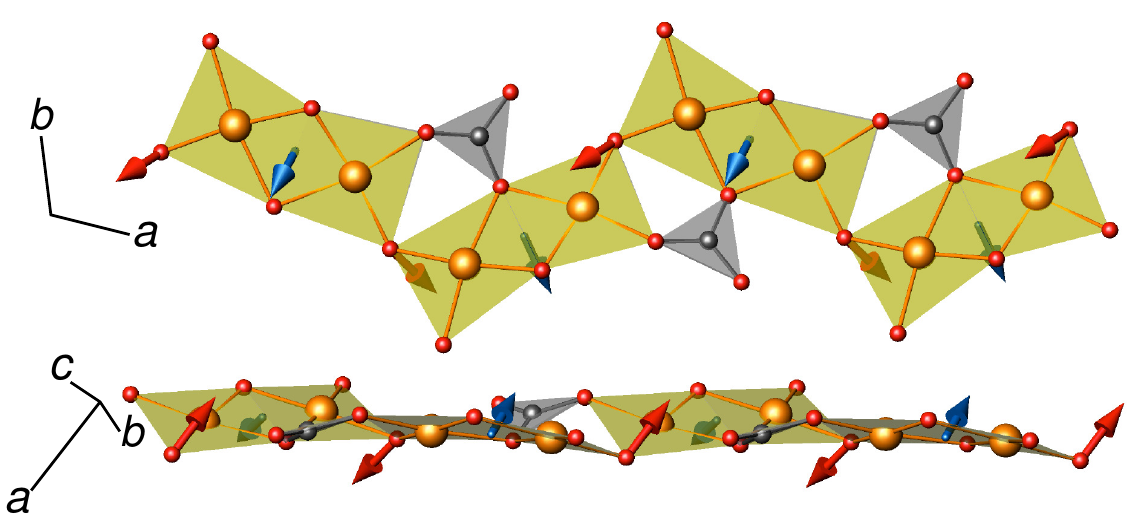}
\caption{\label{F-DM} (Color online) Dzyaloshinskii-Moriya vectors within, $\vec{D}$, and between the dimers, $\vec{D}'$ of the
alternating spin chains of malachite. Note that the neighboring vectors of the
same type ($\vec{D}$ or $\vec{D}'$) are not identical: the signs of the
$x$ and $z$ components alternate along the chain, while the $y$ component retains its sign.}
\end{figure}

To estimate the Dzyaloshinskii-Moriya couplings in malachite, we perform
full-relativistic GGA+$U$ calculations using \texttt{VASP}, and map the
resulting total energies onto a generic bilinear exchange model:

\begin{equation}
    \hat{H} = \sum_{i>j}\sum_{\alpha,\beta}\mathbf{M}_{\alpha,\beta}\hat{S}^{\alpha}_i\hat{S}^{\beta}_j \hskip .05\textwidth \alpha,\beta = x, y, z,
\end{equation}
where $M$ is a 3$\times$3 matrix. Three independent components of its
antisymmetric part define the respective Dzyaloshinskii-Moriya vector
$\Dv_{ij}$:

\begin{equation}
    \hat{H}_{DM} = \sum_{i>j}\mathbf{D}_{ij}\cdot(\mathbf{S}_i\times\mathbf{S}_j).
\end{equation}
For the two dominant couplings, we find $\Dv=[-11.7, -11.3, 10.0]$\,K
and $\Dv'=[-16.3, -7.5, 11.9]$\,K, with the Cu--Cu bond vectors $[-0.266, 0.105, -0.505]$ and $[0.234, -0.181, -0.505]$, respectively. The components are given in the crystallographic coordinate system with the axes $a$, $b$, and $c$. As can be seen in Fig.~\ref{F-DM}, the orientation of $\Dv$ on the
neighboring dimers is different: only the $y$ component is preserved, while the
$x$ and $z$ components of $\Dv$ change sign. As a result, the $\Dv$
vectors on the neighboring dimers are at a skew angle to each other. The same
trend holds for the $\Dv'$ vectors (Fig.~\ref{F-DM}) and emerges naturally from the Moriya rules.\cite{dm2}

A standard estimate for the magnitude of Dzyaloshinskii-Moriya coupling is the
$|\Dv_{ij}|/J_{ij}$ ratio. In malachite, these ratios amount to 0.11 and 0.26 for
$J$ and $J'$, respectively. Since magnetic dimers are formed on the $J$
bonds, the low-energy behavior is governed by $|\Dv|/J$\,=\,0.11.
Qualitatively, any non-zero $D/J$ conforms to the experimental $M(H)$ behavior.
Quantitative estimates can be done by simulating the full microscopic model
with Heisenberg and Dzyaloshinskii-Moriya couplings, and a subsequent averaging
in order to reproduce the behavior of a powder sample. However, such an analysis is
beyond the scope of the present manuscript.

\section{Malachite at high pressures}
\label{sec:highP}

According to our calculations, malachite is one of the rare cases where
structural and magnetic dimers coincide. Other examples are
Cu$_2$As$_2$O$_7$,\cite{cu2as2o7}
SrCu$_2$(BO$_3$)$_2$\cite{SrCu2BO32,SrCu2BO32_P} and
TlCuCl$_3$.\cite{tlcucl3_HP,tlcucl3_P} While the first compound exhibits
magnetic LRO below 10\,K, the latter two remain magnetically disordered and
feature small spin gaps of 34\,K and 7.7\,K, respectively, that can
considerably be affected by pressure. Experimentally, pressure leads to the
closing of the spin gap and LRO. On the model side, this effect implies the
increase in the $\bar{J'}/J$ ratio, where $\bar{J'}$ is a sum of all interdimer
couplings which, in case of malachite, basically include $J'$, $J_{\perp}$ and
the interlayer couplings. 

Different microscopic mechanisms affecting the $\bar{J'}/J$ ratio have been discussed on an empirical level (see Sec.~\ref{sec:discussion}). However, as long as the reliable high-pressure structural data are missing, no analysis of the pressure effects on the microscopic magnetic model of a dimer compound could be performed. Results of a recent DFT-study on the pressure dependence of exchange couplings of CuO\cite{cuo_p} cannot be applied because of the different crystal structure of CuO. 

Therefore, we take advantage of existing high-pressure XRD data for malachite up to 5.17\,GPa\cite{malachiteP} to investigate effects of pressure on the individual exchange couplings. Missing hydrogen positions were obtained by structural optimization within GGA. The transfer integrals $t_{ij}$ and exchange couplings $J_{ij}$ for different pressures together with the evolution of the bridging angles are condensed in Table~\ref{tJP}. An extended table is provided in the supplementary material.\cite{supplement}

\begin{table}[tbp]
\begin{ruledtabular}
\caption{\label{tJP}
Pressure dependence of the magnetic properties and relevant structural
parameters in malachite. ang($J$) gives the evolution of the nearest-neighbor
bridging angles (deg) under pressure (GPa). The subscripts $<$ and $>$ denote the smaller and
larger bridging angles involved in the intradimer coupling $J$. Transfer
integrals $t_{ij}$ (in meV) the total exchange couplings $J_{ij}$ (in K), calculated
with LSDA+$U$ and $U_d$\,=\,8.0\,eV, and the QMC-simulated spin gap $\Delta$
(in K), are given for different pressures. The $t_{ij}$(opt) are obtained with
relaxed atomic positions. For an extended list, including error bars for the bridging angles, see the supplementary material.}
\begin{tabular}{c c c c c c }
pressure  & 0.0 & 1.02  & 2.03 & 3.10 & 5.17 \\ \hline
ang$_<$($J$) & 94.75 & 95.53 & 93.23 & 90.16 & 91.23 \\
ang$_>$($J$) & 106.41 & 105.56 & 105.47 & 107.00 & 105.67 \\
ang($J'$) & 122.13 & 123.19 & 120.95 & 123.33 & 121.62 \\
\\
$t$     & $-143$ & $-135$ & $-127$ & $-114$ & $-111$ \\ 
$t'$ & $-99$ & $-91$   & $-83$  & $-83$  & $-94$ \\
$t_{\perp}$ & $-97$ & $-88$  & $-91$  & $-86$  & $-68$ \\ 
\\
$J$     & 193 & 164   & 159  & 136  & 114 \\ 
$J'$ & 109  & 110   & 105   & 97   & 129 \\
$J_{\perp}$ & 50  & 40    & 25   & 40   & 11 \\ 
\\
$\Delta$ & 77  & 51    & 58   & 27   & 17 \\
\\
$t$(opt) & -- & $-137$ & $-135$ & $-125$ & $-101$ \\
$t'$(opt) & -- & $-93$ & $-96$ & $-88$ & $-75$ \\
$t_{\perp}$(opt) & -- & $-82$ & $-71$ & $-59$ & $-37$ \\
\end{tabular}
\end{ruledtabular}
\end{table}

Regarding the crystal structure, the applied pressure has its strongest effect on the interlayer distance, according to the weak bonding between the structural layers in malachite. Thus, the $a$ and $c$ lattice constants are reduced from 9.502 to 9.114\,\r{A} and 3.240 to 3.057\,\r{A}, respectively, and the enclosed monoclinic angle decreases by about 3.7$^{\circ}$ when increasing the pressure from 0 to 5.17\,GPa. This entails a diminishing Jahn-Teller distortion on the Cu2 site, where the apical Cu--O distance decreases from 2.37\,\r{A} to 2.15\,\r{A} and approaches the in-plane distances of $1.9-2.1$\,\r A. This results in a nearly octahedral coordination, which is highly unfavorable for a $3d^9$ ion. At pressures above 6\,GPa, the system, therefore, undergoes a phase transition to the rosasite structure. Cu2 is then again four-fold coordinated, but with a modified CuO$_4$ plaquette plane lying perpendicular to its orientation in the malachite structure. This transformation involves an abrupt increase in the longest lattice parameter by about $+0.6$\,\r{A}.\cite{malachiteP} The rosasite structure consists of planar chains of edge-sharing CuO$_4$ plaquettes, running along the $c$ axis, linked by Cu1-monomers. The magnetic behavior of this high-pressure phase definitely deserves a closer examination, which is, however, beyond the scope of this work. 

Within the structural layers of malachite, the most prominent effect of pressure is the sizable reduction of the smaller intradimer Cu2--O--Cu1 bridging angle by about 3.5$^{\circ}$ (see Table~\ref{tJP}) which is accompanied by a large increase of the Cu2--O distance by about 0.1\,\r{A}. The larger intradimer angle as well as the Cu--O--Cu interdimer angle of the $J'$ exchange pathway vary both unsystematically by about 2$^{\circ}$, only. 

The effects of the applied pressure on the exchange couplings are most pronounced for the intradimer coupling $J$ which is reduced by more than 40\%. This can be directly related to the evolution of $t$ which, according to the GKA rules, decreases in terms of its absolute value with the decreasing intradimer bridging angle. $J'$ exhibits an unsystematic variation within $\pm20$\,K that cannot be perfectly related to the evolution of the corresponding bridging angle. It seems that also changes of the twisting angle, the angular dependence of the ferromagnetic contribution and may be even more subtle effects determine the pressure dependence of $J'$. The same is true for the interchain coupling $J_{\perp}$: Though one could argue that an increase of the exchange pathway by about 0.1\,\r{A} is responsible for the decreasing coupling strength, it is most likely also effected by distortions of the carbonate group and slightly enhanced buckling of the structural planes. A more detailed analysis of the pressure effects and a comparison with SrCu$_2$(BO$_3$)$_2$ and TlCuCl$_3$ will be given in Sec.~\ref{sec:discussion}.

The LSDA+$U$ estimates for the leading exchange integrals allow for tracing the evolution of the spin gap under pressure. Using QMC, we simulate the field-dependent magnetization, which is zero for a gapped state and non-zero otherwise. QMC simulations cannot be performed for $T$\,=\,0, hence we compute magnetization isotherms at sufficiently low temperature of 0.01\,$J$ (corresponds to 1.9\,K). To correct for finite size effects, we evaluate magnetization for different finite lattices of $N$ spins and extrapolate to the $N\rightarrow\infty$ limit. This way, we obtain the values listed in Table~\ref{tJP}. The general decrease of the spin gap can safely be established, yet the absolute values of the spin gap are not perfectly reliable. For instance, the simulated ambient-pressure spin gap is 77\,K, which is substantially smaller than the experimental value (119\,K). This difference originates from inaccuracies in the values of the leading exchanges: LSDA+$U$ overestimates
$J_{\perp}$ and also $J'$. Thus, LSDA+$U$ values lead to an underestimation of the spin gap at ambient pressure. An additional source of inaccuracies is the restriction to a 2D magnetic model, where a coupling between the magnetic layers is neglected. With increasing pressure these couplings, however, become more effective\cite{supplement} and will further reduce the spin gap. 

Though the overall trend clearly shows a reduction of the spin gap, there is an unexpected sudden increase at 2.03\,GPa (Table~\ref{tJP}). We had a closer look at bond angles and distances but could not find any obvious reason for this behavior, although in general pressure evolution of the bond lengths and distances is somewhat non-monotonic. As structure determination under pressure may be less accurate than in ambient conditions, we made an additional test by performing a full relaxation of all atomic positions of the high-pressure crystal structures (only the lattice parameters were fixed to their experimental values). Exact results of such a relaxation will, to some extent, depend on the exchange-correlation functional, $U_d$ parameter, spin arrangement, and other details of the calculation. Here, we chose GGA+$U$ method, as implemented in \texttt{VASP}, as a suitable reference and considered the AFM spin arrangement. The leading transfer integrals obtained from the relaxed structures are displayed in Table~\ref{tJP}. All $t_{ij}$(opt) show a smooth pressure dependence without any peculiarities at 2.03\,GPa. These results, thus, establish the aforementioned trend for the evolution of the exchange couplings and clearly indicate a substantial decrease of the spin gap under pressure. Such a reduction should be well visible experimentally. A more detailed analysis of the deviations between experimental and relaxed structures as well as a revisiting of the experimental structure of malachite, at least at 2.03\,GPa, are left to future studies.

\section{Discussion}
\label{sec:discussion}

In the present work, we have discussed a microscopic magnetic model for the
famous Cu-mineral malachite. Despite the layered crystal structure, the
magnetization data at ambient pressure can be described by weakly coupled AFM
alternating chains. Intrachain exchange couplings of $J=191$\,K and $J'=86$\,K
open a large spin gap, which is slightly reduced (8\%) to 119\,K by interchain
exchanges conveyed by carbonate groups. The unfrustrated couplings of dimerized
spin chains via polyanions typically facilitate LRO as, e.g., in
Cu$_2$P$_2$O$_7$~\cite{cu2x2o7} and Cu$_2$As$_2$O$_7$.\cite{cu2as2o7}
In contrast, no signs of the LRO have been observed in malachite at least down
to 2\,K. 

\begin{figure}[tb]
\includegraphics[width=8.6cm]{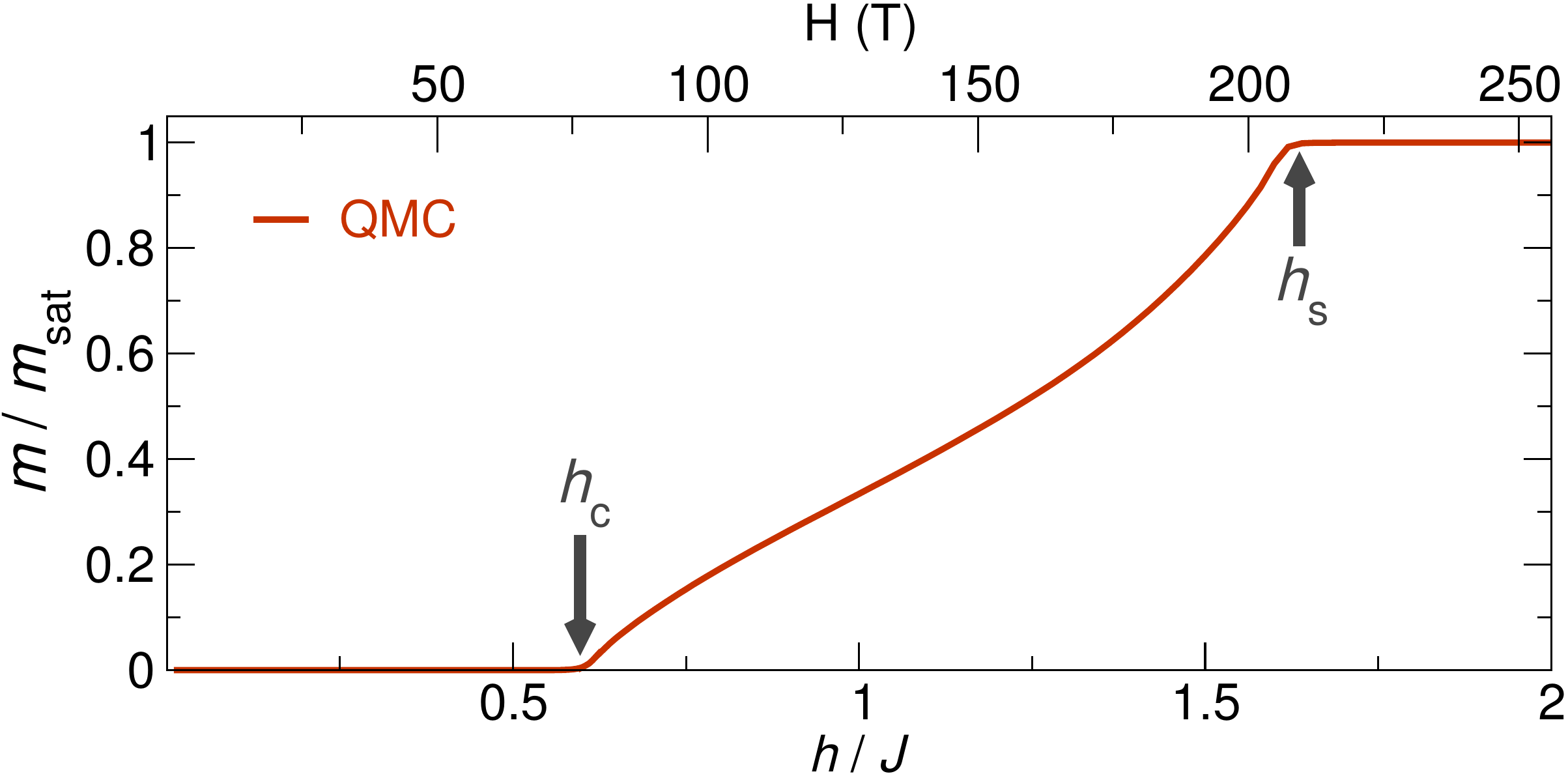}
\caption{\label{F-MH} (Color online) Magnetization isotherm for the
$J-J'-J_{\perp}$ model with $J:J':J_{\perp}$\,=\,1:0.45:0.15, simulated using
quantum Monte Carlo (QMC). The steep increase of $m$ above $h_c$
(closing of the spin gap) and below the saturation field ($h_s$) are
characteristic features of quasi-one-dimensional systems.
}
\end{figure}

To investigate the role of interchain couplings in malachite, we
simulate the magnetization isotherm of the 2D $J-J'-J_{\perp}$ model
using QMC (Fig.~\ref{F-MH}). As expected for a gapped system,
magnetization remains zero up to the critical field $h_{c}$. The $m(h)$
behavior right above $h_{c}$ is ruled by the dimensionality of the
system. Thus, in a 1D model at zero temperature, magnetization behaves
as $m\!\propto\!\sqrt{h^2-\Delta^2}$,\cite{dzhaparidze1978} leading to a
steep increase right above $h_{c}$ (i.e., for $h\!\simeq\!\Delta$), which
eventually transforms to the linear dependence $m\!\propto\!h$ for
$h>\Delta$. Two-dimensional models show a remarkably different
behavior: the $m(h)$ slope is linear, with logarithmic corrections in
the vicinity of $h_c$.\cite{sachdev1994} The $m(h)$ dependence in
Fig.~\ref{F-MH} is closer to the former scenario, indicating a
quasi-one-dimensional behavior.

Another characteristic feature is the transition to the fully polarized
state (saturation). A common feature of one-dimensional magnets is a
steep increase of $m$ right below the saturation field
$h_s$.\cite{schmidt1996} Again, malachite shows a clear resemblance to
the one-dimensional scenario: a pronounced upward bending below $h_s$ is
clearly visible (Fig.~\ref{F-MH}). Thus, we can conclude that the
interchain couplings in malachite play a minor role and this mineral is a
quasi one-dimensional magnet.

A possible explanation for the quasi-1D character of malachite can be found by comparing the spin lattices of different spin-dimer compounds. In malachite, there are two couplings $J_{\perp}$ at each Cu2 site and no interchain couplings at the Cu1 site, thus yielding one interchain coupling per Cu site in average. In Cu$_2$P$_2$O$_7$ and Cu$_2$As$_2$O$_7$, each Cu atom interacts with four Cu atoms from the neighboring chains.\cite{cu2x2o7} Therefore, the tendency of these compounds to the LRO is much stronger compared to malachite.

The DFT results further allow locating the $J$ and $J'$ exchanges in the crystal structure. In a former experimental study,\cite{janod2000} the strongest coupling $J$ has empirically been located between the structural dimers where, according to the GKA rules, the large bridging angle of 122.1$^{\circ}$ should lead to a strong coupling. Our results, however, unambiguously reveal that the intradimer coupling is strongest. Our explanation for this is twofold: First, sizable exchange coupling for an averaged intradimer bridging angle of about 100$^{\circ}$ is indeed not uncommon in Cu-compounds, as we have shown in Sec.~\ref{sec:mag_mod}. Second, we stress that the effect of nonplanar arrangement of neighboring WFs has to be considered in the empirical modelling. The tilting of the interacting CuO$_4$ plaquettes by about 20$^{\circ}$ for $J'$ is responsible for the rather small interdimer coupling. Accordingly, malachite is one of the rare cases where magnetic and structural dimers coincide.

The coinciding magnetic and structural dimers have been previously observed in TlCuCl$_3$ and SrCu$_2$(BO$_3$)$_2$, which attracted large attention because of their small spin gaps of 7.7\,K~\cite{tlcucl3_gap} and 34\,K~\cite{srcubo_gap} and the quantum phase transitions that were observed by applying pressure or magnetic fields.\cite{tlcucl3_HP,SrCu2BO32_P} The possibility of directly observing magnetic ordering processes and quantum phase transitions is of enormous importance for understanding collective quantum phenomena.\cite{tlcucl3_bec}
Which mechanism actually closes/reduces the spin gap under pressure was controversially discussed in the literature. Structural dimers are often considered rigid and unaffected by pressure. Therefore, the reduction of the spin gap was ascribed to the increased interdimer couplings.\cite{srcubo_gap,tlcucl3_gap} On the other hand, the bridging angles for the intradimer coupling $J$ are close to the range of 95--98$^{\circ}$, where a transition from AFM to FM coupling can be expected.\cite{braden} Therefore, the dimers could be very sensitive even to small structural changes. The latter scenario is supported by a work of Johannsen \textit{et al}.~\cite{tlcucl_p} who analyzed susceptibility and magnetostriction data for TlCuCl$_3$. However, these authors set the pressure dependence of the interdimer couplings to zero and used a very simplified magnetic model as a basis for their analysis of pressure effects. 
Though the spin gap of malachite is way larger than in TlCuCl$_3$ and SrCu$_2$(BO$_3$)$_2$, the mechanisms and effects on the exchange couplings induced by pressure should be similar. We thus used existing high-pressure structural data of malachite for the first DFT-based microscopic analysis of pressure effects on exchange couplings in a dimer compound and the typical pressure induced decrease of the spin gap. 

According to our DFT results, external pressure reduces the intradimer coupling $J$ by about 40\%, which can be attributed to the bridging angles decreasing from 94.7$^{\circ}$ to 91.2$^{\circ}$ and from 106.4$^{\circ}$ to 105.7$^{\circ}$, respectively. The interdimer coupling $J'$ varies unsystematically within $\pm20$\,K which is driven by small changes in the bridging angle ($\approx2^{\circ}$) and the bonding distance ($\approx0.03$\,\r{A}).\cite{supplement} The long-range interchain coupling $J_{\perp}$ is reduced from about 50\,K to 12\,K. However, as our QMC simulations for the ambient pressure data have suggested, LSDA+$U$ overestimates $J_{\perp}$ so that this coupling is in fact even smaller and thus of minor importance, in particular at high pressures. Therefore, the evolution of the spin gap is basically driven by the intradimer coupling, which confirms the results of Ref.~\onlinecite{tlcucl_p}. These authors have found d$J$/d$p$\,=\,22\,K/GPa and d$\Delta$/d$p$\,=\,14\,K/GPa for the pressure dependence of the intradimer coupling and the spin gap, respectively. The latter value stems from experimental data, while the former one is derived from a simple magnetic model. In the case of SrCu$_2$(BO$_3$)$_2$, different methods provided different estimates for d$\Delta$/d$p$ in the range 6--11\,K/GPa.\cite{SrCu2BO32_P,SrCu2BO32_sus} For malachite, DFT calculations in combination with QMC simulations supplied the average ratios d$J$/d$p$\,=\,12\,K/GPa and d$\Delta$/d$p$\,=\,11\,K/GPa.\footnote{The 2.03\,GPa data is neglected in this estimate according to the ambiguities in the crystal structure we had found at this pressure.} 
We thus can conclude that the spin gap in dimer compounds is generally reduced by applying pressure. The changes of $\Delta$ and $J$ are thereby similar for the different compounds and, thus, suggest that the same microscopic mechanisms are effective. We have demonstrated that pressure has its main impact on the intradimer coupling, which is crucially responsible for the closing/reduction of the spin gap. However, as our analysis revealed, interdimer couplings in fact cannot $a$ $priori$ be regarded as constant and/or negligible. Thus, in general lots of subtle details have to be taken into account for a quantitative analysis of pressure-induced effects. 

Some of the spin-dimer compounds lack inversion symmetry in the middle of the spin dimer, which gives rise to anisotropic Dzyaloshinskii-Moriya interactions. In SrCu$_2$(BO$_3$)$_2$, these interactions mixing singlet and triplet states are strong enough to invalidate the description of field-induced transition in terms of the Bose-Einstein condensation of magnons.\cite{srcubo_dm} The Dzyaloshinskii-Moriya vectors we estimated for malachite, $|\Dv|/J$\,=\,0.11 and $|\Dv'|/J'$\,=\,0.26, are much stronger than those in SrCu$_2$(BO$_3$)$_2$ ($|\Dv|/J\simeq 0.05$, $|\Dv'|/J'<0.02$).\cite{mazurenko2008} Moreover, $|\Dv'|/J'\gg |\Dv|/J$, which is opposite to the situation in SrCu$_2$(BO$_3$)$_2$. Considering the large DM anisotropy in malachite, we suggest that experimental studies of the magnetic excitation spectrum by electron spin resonance and/or inelastic neutron scattering could be insightful. For example, we envisage a peculiar splitting of the triplet band, similar to the recent observation of two nearly parallel bands in the frustrated-spin-ladder compound BiCu$_2$PO$_6$.\cite{tsirlin2010b,bicupo} Experimental studies on the magnetism of malachite under pressure should be interesting as well. Our work provides solid microscopic basis for such studies.\\

\section{Summary}
\label{sec:summary}

In summary, we have performed a combined theoretical and experimental study on the famous Cu-mineral malachite at ambient pressure and have made predictions for the evolution of its spin gap for pressures up to 5.17\,GPa. For the magnetic modelling of the high-pressure structures, we first determined the hydrogen-positions, missing in the presently available XRD-data, by structural optimizations within DFT. The reliability of this method has been tested on ambient pressure malachite for which accurate neutron data exist. The results are in excellent agreement with the experimental data, so that we propose the determination of hydrogen positions by DFT as a highly valuable, fast and cheap alternative to experiments. 

The magnetic structure of malachite at ambient pressure is well described by the model of alternating antiferromagnetic chains with the couplings $J=191$\,K and $J'=86$\,K. Interchain couplings slightly reduce the resulting spin gap by 8\% to 119\,K. The evolution of the exchange couplings and the spin gap under pressure has been investigated by DFT calculations and QMC simulations. The results have been compared with the dimer compounds TlCuCl$_3$ and SrCu$_2$(BO$_3$)$_2$ for which different mechanisms for the closing of the spin gap have been proposed. In this study, we have explicitly demonstrated that the reduced intradimer coupling is the driving force for closing the spin gap under pressure. Further, Dzyaloshinskii-Moriya interactions were estimated and assigned to be responsible for the linear increase in the magnetization at low fields.

\acknowledgments
We are grateful to Gudrun Auffermann for her kind help with the chemical analysis of small mineral samples. We also acknowledge the experimental support
by Yurii Prots and Horst Borrmann (laboratory XRD), Caroline Curfs (ID31), and the provision of the ID31 beamtime by the ESRF. We are grateful to Yuri Skourskii for high-field magnetization measurements. AT is thankful to Ivan Bushmarinov for his instructive comments on the experimental determination of the hydrogen position. We would like to thank the Department of Materials Research and Physics of the Salzburg University for providing the natural sample of malachite from their mineralogical collection (inventory number 15002). AT and OJ were supported by the Mobilitas program of the ESF (grant numbers MTT77 and MJD447). SL acknowledges the funding from the Austrian Fonds zur F\"orderung der wissenschaftlichen Forschung (FWF) via a Schr\"odinger fellowship (J3247-N16).


%

\clearpage

\begin{table*}[h]
\begin{tabular}{c}
\huge{\texttt{Supporting Material}} \\
\end{tabular}
\end{table*}

\begin{widetext}

\begin{figure}[H]
\includegraphics[width=17.1cm]{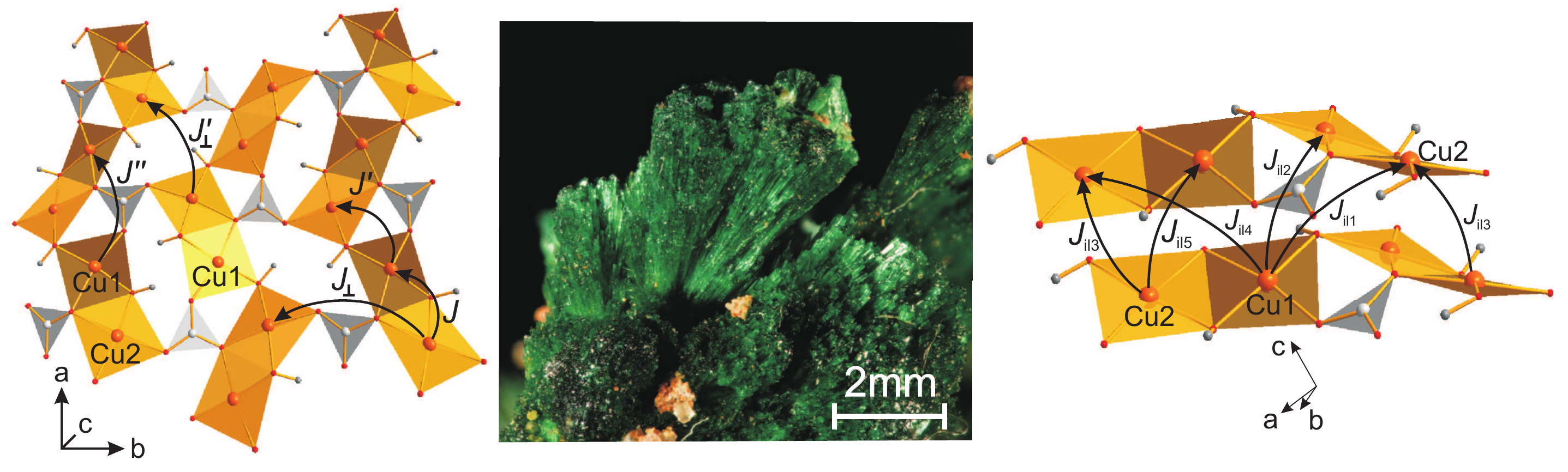}
\caption{\label{structure}
(Color online) The crystal structure of malachite. The left panel shows a single layer consisting of dimer-chains linked by CO$_3$ groups (gray). The relation between the layers is depicted in the right pane. Arrows indicate the leading hopping pathways. The central panel shows the natural sample of malachite.}
\end{figure}

\begin{table}[!h]
\begin{ruledtabular}
\caption{\label{tJ} 
Transfer integrals $t_{ij}$ (in meV) and the AFM exchange contributions $J^{\text{AFM}}_{ij}=4t^2_{ij}/U_{\text{eff}}$ (in K) where $U_{\text{eff}}$\,=\,4.5\,eV. The total exchange couplings $J_{ij}$ are calculated with the LSDA+$U$ method and the parameters $U_d=8.0\pm1.0$\,eV, $J_d=1.0$\,eV. }
\begin{tabular}{c c c c c}
   &  Cu-Cu distance (\r{A}) & $t_{ij}$ & $J^{AFM}_{i}$ & $J_{ij}$  \\ \hline
$J$     & 3.06 & -143 & 211 	& $193\pm40$  	\\ 
$J'$ & 3.34 & -99 	& 101 	& $109\pm20$  	\\
$J_{\perp}$ & 6.03 & -97 	& 96 	& $50\pm10$  	\\ 
$J''$ & 5.40 & 53 	& 29 	& $4.1\pm0.5$ 	\\
$J_{\perp}'$ & 5.38 & 38 	& 15 	&   			\\
$J_{il1}$ & 4.84 & -41	& 17 	&   			\\
$J_{il2}$ & 3.67 & 37 	& 14 	& $-0.8\pm0.5$  	\\
$J_{il3}$ & 3.24 & 71 	& 52 	& $-8\pm1$  \\
$J_{il4}$ & 5.29 & 29 	& 9 	&   			\\
$J_{il5}$ & 3.43 & 30 	& 9 	&   			\\
\end{tabular}
\end{ruledtabular}
\end{table}

\begin{table}[!h]
\begin{ruledtabular} 
\caption{\label{H_rot} 
Bonding distances (\r{A}) and bridging angles (deg) in the experimental high-pressure structures of malachite.
(ap) denotes the apical oxygen atoms.
}
\begin{tabular}{c c c c c c c}
		& 		& 	\multicolumn{5}{c}{Pressure (GPa)}  \\ 
\multicolumn{2}{c}{distances} & 0.00 & 1.02  &  2.03  &  3.10  &  5.17  \\ \hline
   Cu1	&	O4	&	1.898(1) &	1.831(75) &	1.890(47) &	1.845(23) &	1.860(23)	\\
		&	O5	&	1.911(1) &	1.914(31) &	1.913(52) &	1.884(21) &	1.904(71)	\\
		&	O1	&	1.996(1) &	1.985(20) &	1.976(50) &	1.962(15) &	1.964(15)	\\
		&	O2	&	2.053(1) &	2.005(16) &	2.046(51) &	2.069(13) &	2.050(14)	\\
		& O1(ap)&	2.510(1) &	2.461(19) &	2.407(57) &	2.373(15) &	2.325(15)	\\
		& O2(ap)&	2.639(1) &	2.566(17) &	2.552(42) &	2.522(13) &	2.414(15)	\\
		& Cu2($J$)	&	3.063(1)	&	3.056(5) &	3.053(13) &	3.058(4) &	3.055(46)	\\
		& Cu2($J'$)&	3.340(1)	&	3.328(7) &	3.329(5)  &	3.326(6) &	3.324(61)	\\
   Cu2	&	O5	&	1.915(1) &	1.924(12) &	1.924(53) &	1.920(11) &	1.930(70)	\\
		&	O4	&	1.918(1) &	1.952(80) &	1.935(55) &	1.952(13) &	1.948(13)	\\
		&	O3	&	2.049(1) &	2.033(24) &	2.010(34) &	2.069(40) &	2.099(20)	\\
		&	O2	&	2.110(1) &	2.122(23) &	2.154(139) & 2.246(17) & 2.222(20)	\\
		& O5(ap)&	2.369(1) &	2.339(13) &	2.330(58) &	2.272(13) &	2.183(75)	\\
		& O4(ap)&	2.373(1) &	2.321(94) &	2.298(63) &	2.234(13) &	2.153(14)	\\
\\
\multicolumn{2}{c}{angles} &  &   &   &   &   \\ 	\hline
$J$	    &	\mbox{Cu1--O2--Cu2}	& 94.75(4)  & 95.53(68)   &	93.23(42)  & 90.16(47)	& 91.23(53) \\
		&	\mbox{Cu1--O5--Cu2}	& 106.42(5) & 105.56(59)  &	105.47(57) & 107.00(57)	& 105.67(112) \\
$J'$	&	\mbox{Cu1--O4--Cu2}	& 122.13(5) & 123.19(130) &	120.95(92) & 123.33(73)	& 121.62(75) \\
\end{tabular}
\end{ruledtabular}
\end{table}

\begin{table}[!h]
\begin{ruledtabular}
\caption{\label{tJP} 
Transfer integrals $t_{ij}$ (in meV) the total exchange couplings $J_{ij}$ (in K),
calculated with LSDA+$U$ and $U_d=8$\,eV, $J_d=1$\,eV for different pressures (in GPa).}
\begin{tabular}{c c c c c c }
pressure  & 0.0 & 1.02  & 2.03 & 3.10 & 5.17 \\ \hline
$t$     & -143 & -135 & -127 & -114 & -111 \\ 
$t'$ & -99 & -91   & -83  & -83  & -94 \\
$t_{\perp}$ & -97 & -88   & -91  & -86  & -68 \\ 
$t''$ & 53 & 50     & 49   & 48   & 44  \\
$t_{\perp}''$ & 40 & 36     & 52   & 36   & 30  \\
$t_{il1}$ & -41 & -44   & -41  & -41  & -43 \\
$t_{il2}$ & 37 & 44     & 35   &  45  & 55  \\
$t_{il3}$ & 71 & 71     & 64   & 70   & 84  \\
$t_{il4}$ & 29 & 31     & 27   & 28   & 36  \\
$t_{il5}$ & 30 & 30     & 27   &  30  & 39  \\
\\
$J$     & 193 & 164   & 145  & 136  & 114 \\ 
$J'$ & 109  & 110   & 70   & 97   & 129 \\
$J_{\perp}$ & 50  & 40    & 49   & 40   & 11 \\ 
$J''$ & 2  	& 0.5 	& 10   & 9	  & 4 \\
$J_{il2}$ & -1  & 6    & -1.5    & 6   & 19 \\
$J_{il3}$ & -1& -2.5  & -4 & -6 & -5 \\
\end{tabular}
\end{ruledtabular}
\end{table}

\begin{table}[!h]
\caption{
Refined atomic positions (in fractions of lattice parameters) and isotropic atomic displacement parameters $U_{\text{iso}}$ (in $10^{-2}$~\r A$^2$) for the malachite structure at 80~K (first row) and at room temperature (second row). All atoms are in the general position $4e$ of the space group $P2_1/a$. Lattice parameters are as follows: $a=9.47697(3)$~\r A, $b=11.93401(5)$~\r A, $c=3.22835(1)$~\r A, $\beta=98.5322(3)^{\circ}$ at 80~K ($R_I=0.028$) and $a=9.49454(3)$~\r A, $b=11.95662(4)$~\r A, $c=3.24425(1)$~\r A, $\beta=98.7742(2)^{\circ}$ at room temperature ($R_I=0.046$). The $U_{\text{iso}}$ of oxygen atoms were refined as a single parameter. Hydrogen positions were not refined. All standard deviations are from the Rietveld refinement, only.
}
\begin{minipage}{15cm}
\begin{ruledtabular}
\begin{tabular}{ccccc}
  Atom & $x/a$      & $y/b$      & $z/c$     & $U_{\text{iso}}$ \\
  Cu1  & 0.49904(9) & 0.28809(6) & 0.8875(3) & 0.36(2)          \\
       & 0.49806(8) & 0.28785(6) & 0.8932(2) & 0.73(2)          \\
  Cu2  & 0.23260(8) & 0.39353(7) & 0.3829(3) & 0.30(2)          \\
       & 0.23236(7) & 0.39293(5) & 0.3888(2) & 0.59(2)          \\
  O1   & 0.1329(4)  & 0.1350(3)  & 0.337(1)  & 0.10(4)          \\
       & 0.1340(3)  & 0.1349(3)  & 0.332(1)  & 0.41(4)          \\
  O2   & 0.3315(4)  & 0.2359(3)  & 0.447(1)  & 0.10(4)          \\
       & 0.3295(3)  & 0.2361(3)  & 0.451(1)  & 0.41(4)          \\
  O3   & 0.3334(4)  & 0.0569(3)  & 0.629(1)  & 0.10(4)          \\
       & 0.3344(3)  & 0.0543(2)  & 0.628(1)  & 0.41(4)          \\
  O4   & 0.0956(4)  & 0.3513(3)  & 0.911(1)  & 0.10(4)          \\
       & 0.0948(3)  & 0.3498(3)  & 0.919(1)  & 0.41(4)          \\
  O5   & 0.3781(4)  & 0.4164(3)  & 0.857(1)  & 0.10(4)          \\
       & 0.3785(3)  & 0.4154(2)  & 0.863(1)  & 0.41(4)          \\
  C    & 0.2660(6)  & 0.1405(5)  & 0.468(2)  & 0.3(1)           \\
       & 0.2667(6)  & 0.1378(4)  & 0.473(2)  & 1.1(2)           \\
\end{tabular}
\end{ruledtabular}
\end{minipage}
\end{table}

\begin{figure}[H]
\includegraphics{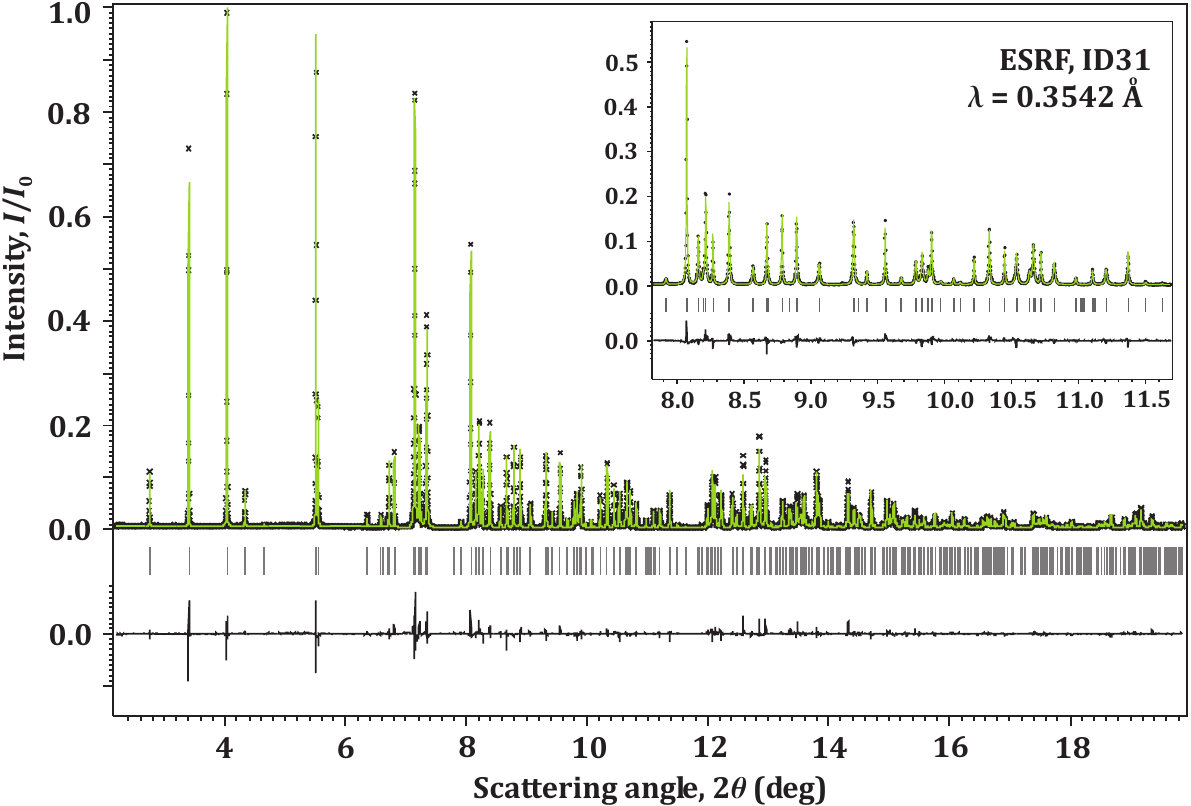}
\caption{
Structure refinement from the synchrotron powder data at 80~K. Ticks denote reflection positions.
}
\end{figure}

\bigskip

\end{widetext}

\end{document}